\documentclass[trackchanges,twocolumn]{aastex701}

\newcommand\msun{M_\odot}

\usepackage{comment}
\usepackage{multirow}
\usepackage[flushleft]{threeparttable}
\usepackage{graphicx}
\usepackage{amsmath}
\usepackage{breqn}
\usepackage{enumitem}
\shorttitle{UV continuum and Balmer lines in LRDs}
\shortauthors{Asada et al.}
\graphicspath{{./}{figures/}}

\begin{document}

\title{Origins of the UV continuum and Balmer emission lines in Little Red Dots:\\
observational validation of dense gas envelope models enshrouding the AGN}

\author[orcid=0000-0003-3983-5438,sname=Asada,gname=Yoshihisa]{Yoshihisa Asada}
\altaffiliation{Dunlap Fellow}
\affiliation{Dunlap Institute for Astronomy and Astrophysics, 50 St. George Street, Toronto, ON M5S 3H4, Canada}
\email[show]{yoshi.asada@utoronto.ca}
\correspondingauthor{Yoshihisa Asada}

\author[orcid=0000-0000-0000-0001,sname='Inayoshi']{Kohei Inayoshi}
\affiliation{Kavli Institute for Astronomy and Astrophysics, Peking University, Beijing 100871, China}
\email[show]{inayoshi@pku.edu.cn}

\author[orcid=0000-0001-7232-5355,sname=Fei,gname=Qinyue]{Qinyue Fei}
\affiliation{David A. Dunlap Department of Astronomy and Astrophysics, University of Toronto, 50 St. George Street, Toronto, ON M5S 3H4, Canada}
\email{qyfei.astro@gmail.com}  

\author[orcid=0000-0001-7201-5066,sname=Fujimoto,gname=Seiji]{Seiji Fujimoto}
\affiliation{David A. Dunlap Department of Astronomy and Astrophysics, University of Toronto, 50 St. George Street, Toronto, ON M5S 3H4, Canada}
\affiliation{Dunlap Institute for Astronomy and Astrophysics, 50 St. George Street, Toronto, ON M5S 3H4, Canada}
\email{seiji.fujimoto@utoronto.ca}

\author[orcid=0000-0002-4201-7367,gname=Chris, sname=Willott]{Chris J. Willott}
\affiliation{NRC Herzberg, 5071 West Saanich Rd, Victoria, BC V9E 2E7, Canada}
\email{chris.willott@nrc.ca}  

\begin{abstract}
We present a statistical study on the origins of the UV continuum and narrow/broad emission lines in little red dots (LRDs), presumably involving active galactic nuclei (AGNs).
Leveraging all archived JWST/NIRSpec data, we build a sample of 27 spectroscopically-confirmed LRDs at $5<z_{\rm spec}<7.2$, by requiring broad H$\alpha$ emission, blue UV colors, V-shaped continua, and compact morphologies.
We define a control sample of 7 blue, compact, broad-line AGNs without red optical continua (hereafter little {\it blue} dots; LBDs), and examine correlations between rest UV and the narrow/broad H$\alpha$ luminosities in these populations.
In LRDs, both narrow and broad H$\alpha$ components are tightly correlated with the UV continuum, and the luminosity ratios are consistent with those in young starburst galaxies.
In contrast, the UV to {\it broad} H$\alpha$ ratios in LBDs closely match local unobscured AGNs and are statistically different from LRDs.
The Ly$\alpha$ occurrence rates and strengths do not differ between LRDs and LBDs and are comparable to normal star-forming galaxies.
These results are consistent with a scenario where the central BH in LRDs is enshrouded by a dense opaque gas envelope -- in this model, the UV continuum as well as narrow and even broad H$\alpha$ emissions are not powered by AGNs but predominantly by young massive stars surrounding the envelope, while the envelope radiates as a $\sim 5000$ K blackbody.
As the envelope dissipates, direct AGN emission can emerge, potentially transforming LRDs into LBDs and marking the end of a short-lived phase of rapid black hole growth.
\end{abstract}



\section{Introduction}\label{sec:intro}


A newly discovered population of broad Balmer line emitters known as little red dots (LRDs; \citealt{Matthee_2024}), presumably involving active galactic nuclei (AGNs), is one of 
the most intriguing findings from the James Webb Space Telescope (JWST).
LRDs predominantly appear at high redshifts ($z\sim 4-8$) and rapidly decline in abundance toward lower redshifts \citep[e.g.,][]{Kocevski2025ApJ,Ma_2025,Zhuang_2025}.
Their unique spectral and morphological properties set them apart from both typical low-redshift AGNs and luminous quasars
\citep[e.g.,][]{Greene_2024,Labbe_2024b,Furtak2024Natur,Akins_2025,Hviding2025AAP,Maiolino_2025_Xray,Ji_2025}.
LRDs may represent a population of black holes (BHs) with masses $M_{\rm BH} \simeq 10^{5}–10^{7}~\msun$, 
caught in a transient phase of rapid growth from seed BHs to supermassive regimes \citep{Inayoshi_2025a}.

The presence of characteristic Balmer-transition signatures in LRDs, including prominent absorption superimposed on broad Balmer emission lines,
unusually large Balmer decrements, deep Balmer breaks, large equivalent widths of broad H$\alpha$ emission,  and exponential line profiles of the broad line components suggest that LRDs are AGNs embedded 
in dense gas, more specifically broad line regions (BLRs) residing within dense gaseous structures
\citep[e.g.,][]{Juodzbalis_2024, Lin_2024,Inayoshi_Maiolino_2025,Ji_2025,deGraaff_2025b,Naidu_2025,Rusakov2025arXiv}.
When such dense gas clumps in BLRs become optically thick and act as Balmer line and continuum emitters,
their thermal emission yields an optically red continuum spectrum by the Wien tail of a blackbody spectrum with an effective temperature of $T_{\rm eff}\simeq 5000~{\rm K}$,
without dust reddening \citep{Inayoshi_2025b,Kido_2025,Liu_2025b,Lin_2025c,Begelman_2025}. 
This interpretation is novel and critical, as JWST/MIRI photometry disfavors significant rest-frame near- to mid-infrared emission from AGN-heated hot dust \citep{Akins_2025,Casey_2025,K.Chen_2025}.
As additional consequences, dense nuclear gas can efficiently cool X-ray emitting hot coronae via Compton scattering \citep{Yue_2024,Maiolino_2025_Xray} and dissipates 
the power of nascent jets within the gaseous envelope \citep{Mazzolari_2024,Gloudemans_2025}.

\begin{figure*}[t]
\centering
\includegraphics[width=0.9\textwidth]{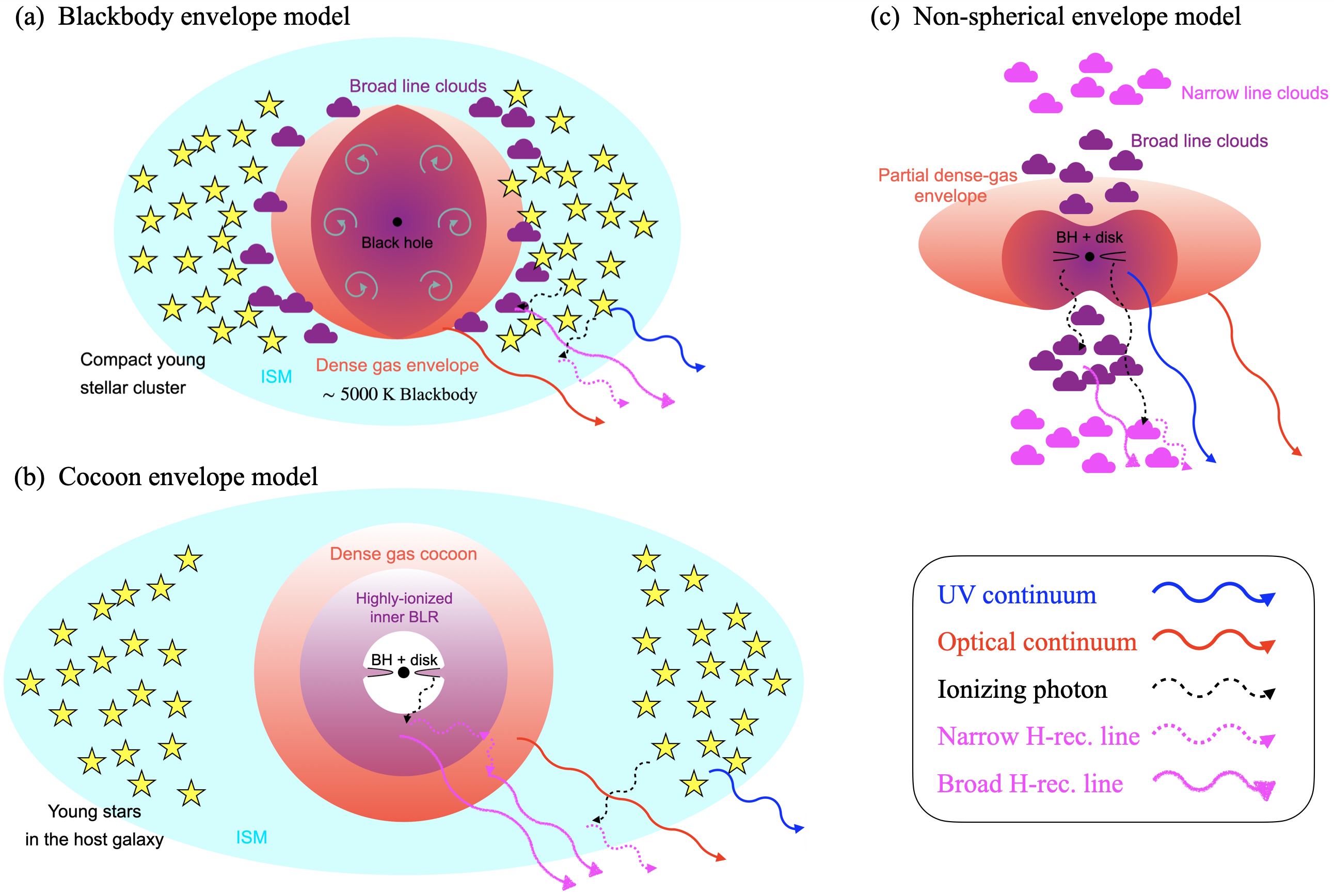}
\caption{Schematic pictures of AGN+dense gas envelope models (also known as BH* models) for LRDs proposed in literature.
(a) \citet{Inayoshi2025arXiv} proposed a model where the central BH is enshrouded by thick gas envelope. Since its optically thickness, no radiation from the AGN accretion can be escaped, and the envelope is thermalized and observed as a blackbody. In this model, all UV light, narrow lines, and broad lines are powered by young massive stars surrounding the envelope, and the rest optical continuum is explained by the thermal emission.
(b) the cocoon envelope model originally proposed by \citet{Naidu_2025} has a layered cocoon around the central BH. The broad lines are originated from the inner layer, which is illuminated by the accretion disk, and leak out after multiple scatterings through the outer layer. The cocoon is optically thick for UV continuum, and the UV light and narrow line in LRDs needs to be emitted from host galaxies. The rest optical continuum in this model originated either from the Balmer limit photon absorption by the cocoon or from the thermal emission of the cocoon heated by the AGN accretion.
(c) when the dense envelope around the central BH is not spherical and is not optically thick towards the polar axis \citep[e.g.,][]{Lin_2025c}, all UV light, narrow, and broad lines can be from the AGN accretion. This is similar to the standard AGN picture, but it lacks the dust torus and instead has a non-spherical envelope which emits blackbody thermal emission in the rest optical.
}
\vspace{5mm}
\label{fig:models}
\end{figure*}

This ``AGN+dense gas" scenario, which has been proposed in various forms in the literature (e.g., envelops, clumps, or BH*; see below), successfully explains many of the puzzling properties of LRDs, including the red optical continuum produced by gaseous thermal emission and selective absorption near the Balmer limit.
However, it faces an important question regarding the origin of the observed UV continuum and broad-line emissions \citep[see also][]{Inayoshi_Ho_2025}.
If the gas envelope fully obscures the UV radiation emitted from the accretion disk, the UV continuum of LRDs should arise from external components,
such as compact young stellar populations.
In this case, depending on the gas clumpiness and the optical depths to line transitions, the BLRs can be powered either by ionizing radiation from the the stellar population \citep[Model A;][]{Inayoshi2025arXiv}
or by radiation reprocessed from the embedded AGN that leaks out after multiple scattering \citep[Model B;][]{Naidu_2025}.
Model (A) requires relatively intense star formation to supply sufficient ionizing photons to sustain the BLRs, whereas Model (B) demands a specific density and clumpiness structure 
in which the envelope is opaque to the UV continuum but allows Balmer lines to escape.
Alternatively, if UV radiation produced from the accretion disk can escape preferentially toward the polar directions owing to rotation or anisotropic gas geometry,
these photons could directly photoionize the BLRs as in the standard AGN picture \citep[Model C;][]{Lin_2025c}.

Each model predicts different observational signatures in the UV continuum and BLR emission (Figure~\ref{fig:models}).
One key diagnostic is the relationship between these luminosities.
In Models (A) and (C), both the UV continuum and broad-line emission are powered by a single radiation source:
stellar ionization photons and the central accretion disk, respectively.
As a result, a tight correlation between the UV and broad-line luminosities is expected, analogous to that observed in 
star-forming galaxies (Model A) and in normal AGNs (Model C), respectively.
In contrast, Model (B) invokes two spatially and physically disconnected radiation sources: the UV continuum is dominated by stellar emission and the BLRs are powered by 
the embedded accretion disk. In this case, the connection between the UV continuum and BL emission is indirect, and any correlation between their luminosities is expected to be weak or absent.

In this paper, we utilize all archived data taken with JWST to date and make a large statistical sample of spectroscopically-confirmed LRDs to examine these envelope models. We also make a control sample of normal unobscured broad-line (Type 1) AGNs at the same redshift range, which do not show V-shaped continuum spectra but instead have blue UV-to-optical color and broad Balmer emission lines, to characterize typical AGN spectra.
Using the spectroscopic samples of LRDs and non-LRD AGNs, we measure their rest UV continuum and Balmer line emissions, and compare the energy balance between the two luminosity components.
This investigation enables us to anchor their energy sources with and without gaseous envelopes surrounding the nuclear BHs.
Throughout the paper, we assume a flat $\Lambda$-CDM cosmology with $\Omega_{\rm m}=0.3$, $\Omega_{\Lambda}=0.7$, and $H_0=70\ {\rm km\ s^{-1}\ Mpc^{-1}}$.

\section{Data and sample selection}\label{sec:data}
We utilize all JWST/NIRSpec Prism spectra available on the DAWN JWST Archive \citep[DJA;][]{Brammer2025zndo}.
Spectra on the DJA version 4.4 are reduced with \texttt{msaexp} \citep{Brammer2022zndo}, following \citet{deGraaff2025AAP} and \citet{Heintz2024Sci}, and the 1D spectra are extracted using the optimal aperture method \citep{Horne1986PASP}.
We only used \texttt{grade=3} NIRSpec/Prism spectra without missing data due to the detector gap from rest far-UV to optical bands.
We focus on the NIRSpec/Prism sample, because one of our goals is to spectroscopically measure the rest UV continuum colors and luminosity. Requiring NIRSpec grating observations would enable us to capture more fainter and/or narrower broad line components, but it would make our sample much more smaller and heterogeneous.
This selection includes 80,367 spectra in total.


We first select broad H$\alpha$ line emitters at $5<z_{\rm spec}<7.2$ with rest UV continuum detection.
Specifically, we use the DJA outputs for the initial selection of sources that meet:
\begin{itemize}
    \item 5 $<$ \texttt{z_best} $< 7.2$,
    \item S/N(H$\alpha$) $> 30$,
    \item the S/N of the synthesized flux of the Prism spectrum in the rest-UV filter is greater than 3,
\end{itemize}
where \texttt{z_best} is the best redshift estimate from the spectrum given as a DJA product.
This initial cut selects 383 H$\alpha$ emitters at $5<z_{\rm spec}<7.2$.
The redshift range is determined so that both the rest UV continuum and H$\alpha$ line can be observed with NIRSpec/Prism.
For each selected H$\alpha$ emitter, we fit the observed H$\alpha$ line twice: with a single unresolved line profile and with two-component (broad+narrow) line profiles. We then compute the least chi-square values of these fits, and select broad H$\alpha$ emitters by requiring $\chi^2_{\rm unres}/\chi^2_{\rm two-comp} > 1.6$.
We finally removed duplicated sources that were observed in multiple programs and registered on DJA as different sources.
We obtain a sample of 50 broad H$\alpha$ emitters.

\begin{figure*}[t]
\centering
\includegraphics[width=0.45\textwidth]{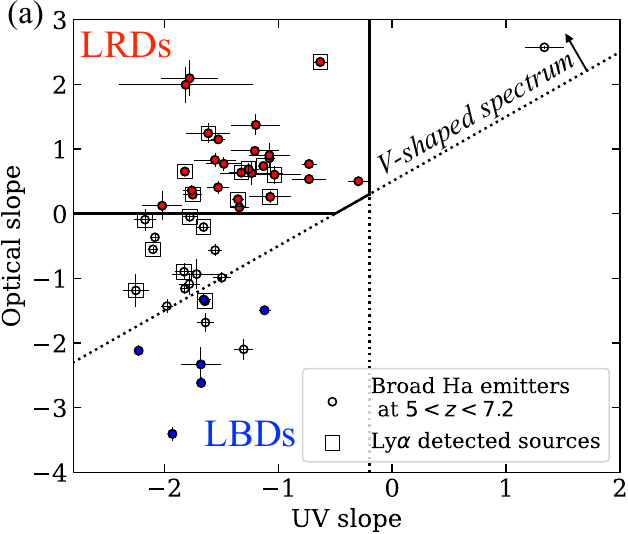}\vspace{2mm}
\includegraphics[width=0.95\textwidth]{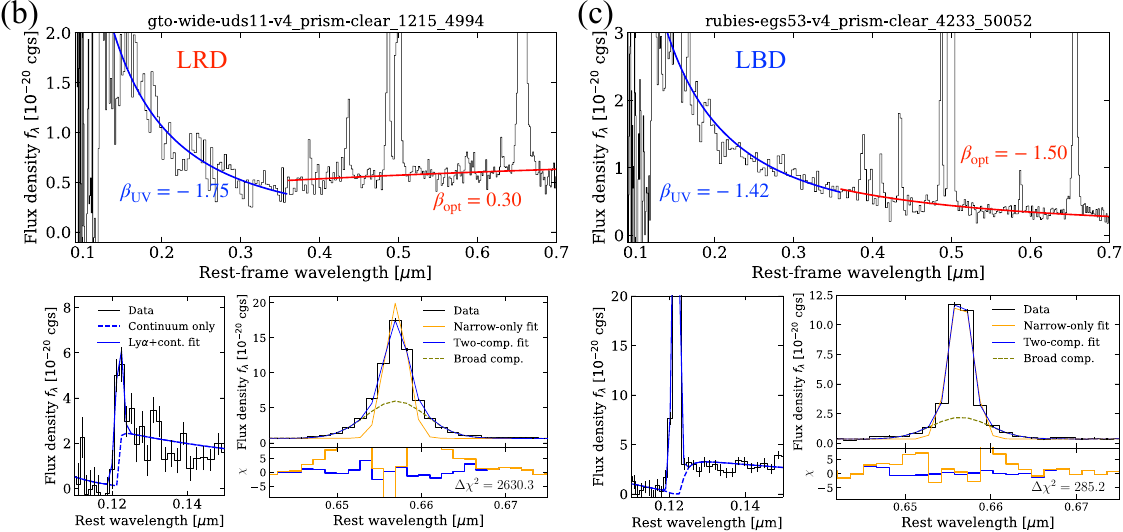}
\caption{Sample selection in this work.
(a) we select LRDs with V-shaped continua from spectroscopically-confirmed broad H$\alpha$ line emitters at $5<z_{\rm spec}<7.2$ in the $\beta_{\rm opt}$-$\beta_{\rm UV}$ diagram.
Thick black lines define the color window for LRDs. As a control sample, we also select blue normal AGNs without the V-shaped continuum in this color-color diagram. 
(b) an example of sample LRDs, which taken as part of the GTO-Wide program \citep{Maseda2024AAP}. The top panel presents the NIRSpec/Prism spectrum, and the bottom panel shows the Ly$\alpha$ (left) and H$\alpha$ (right) line profile. For the Ly$\alpha$ line, the best-fit line profile is shown with the blue solid curve, while the continuum-only fit model is with the dashed curve. For the H$\alpha$ line, the best-fit line profiles with narrow, unresolved component fit (orange) and broad+narrow components fit (blue) are shown, and the noise-normalized residual for each case is presented in the lower sub-panel.
(c) same as panel (b) but for an example of the LBDs, taken as part of the RUBIES program \citep{deGraaff2025AAP}.
}
\vspace{5mm}
\label{fig:sample}
\end{figure*}

We next select LRDs among the broad line emitters based on their colors and compactness. 
Following \citet{Hviding2025AAP}, we define the V-shaped LRD colors as:
\begin{enumerate}
    \item $\beta_{\rm UV}<-0.2$,
    \item $\beta_{\rm opt}>0$,
    \item $\beta_{\rm opt} - \beta_{\rm UV} > 0.5$,
\end{enumerate}
where $\beta_{\rm UV}$ and $\beta_{\rm opt}$ are rest UV and optical slopes measured from the spectra at $1250\leq \lambda_{\rm UV}/{\rm \AA} \leq 3600$ and $3600\leq \lambda_{\rm opt}/{\rm \AA} \leq 7000$, respectively, with masking out wavelength ranges where strong emission lines are expected.
Figure~\ref{fig:sample} (a) presents the $\beta_{\rm opt}-\beta_{\rm UV}$ distribution of the broad H$\alpha$ emitters at $5<z_{\rm spec}<7.2$, and LRDs are marked by red in this diagram.
We finally select compact sources satisfying $f_{\rm F444W}(0.\!\!^{\prime\prime}35)/f_{\rm F444W}(0.\!\!^{\prime\prime}18)<1.7$.
We obtain a sample of 27 LRDs in total, including observations from CANUCS \citep{Sarrouh2025arXiv}, CAPERS (GO-6368; PI: Dickinson), CEERS \citep{Finkelstein2025ApJ}, JADES \citep{Curtis-Lake2025,Scholtz2025arXiv}, NEXUS \citep{Shen2024arXiv}, RUBIES \citep{deGraaff2025AAP}, UNCOVER \citep{Bezanson2024ApJ}, GTO-Wide \citep{Maseda2024AAP}, and other small programs (GO-2198; PI: Barrufet and DD-6585; PI: Coulter).
Figure~\ref{fig:sample} (b) shows an example of LRDs.
We note that extreme cases of LRDs like {\it MoM-BH*-1} \citep{Naidu_2025} or {\it The Cliff} \citep{deGraaff_2025b} could be missed due to the $\beta_{\rm UV}$ criterion, since these extreme ones can sometimes have redder UV colors. However, relaxing the $\beta_{\rm UV}$ criterion could let dusty AGNs/starburst galaxies in the sample. Indeed, there is one galaxy that could pass our criteria when the $\beta_{\rm UV}$ criterion is removed, but the spectrum looks largely different from other LRDs with having very red continuum at all wavelength range up to $\lambda_{\rm rest}\sim8000$ \AA.

We also make a control sample of broad H$\alpha$ emitters without the V-shaped spectrum.
The V-shaped spectrum is a unique feature of LRDs, potentially originating from the dense-gas envelope enshrouding the AGN, and selections without the V-shaped color criteria of LRDs should thus make a sample of typical Type 1 AGNs just without the envelope at the same redshift range and luminosity.
We note that the physical origin of this V-shaped spectrum is under debate, and recent studies on a gravitationally-lensed LRD \citep[A383-LRD1;][]{Golubchik2025arXiv,Baggen2025arXiv} showed the V-shaped spectra in LRDs could be due to the superposition of two clumps with different colors. However, the statistical comparison between LRDs and this control sample itself is not affected by the physical origin assumption.
We make the control sample based on the color-color criteria as
\begin{enumerate}
    \item $\beta_{\rm UV}<-0.2$,
    \item $\beta_{\rm opt} - \beta_{\rm UV} < 0.5$,
\end{enumerate}
and the same compactness criterion, among the broad H$\alpha$ emitters (Figure~\ref{fig:sample} (a) blue).
Hereafter, we refer to this control sample as {\it Little Blue Dots} (LBDs), and use it to probe the effect of enshrouding envelopes in LRDs.
We obtain a sample of 7 LBDs, including observations by \citet{Abdurro'uf2024ApJ} and \citet{Wang2024ApJ} in addition to those for LRDs.
An LBD example is shown in Figure~\ref{fig:sample} (c).

For all LRDs and LBDs, we measure the UV luminosity and broad+narrow H$\alpha$ line luminosity from the NIRSpec/Prism spectrum.
We first fit the continuum around the line with a power-law function, and fit a two-component Gaussian profile on the observed H$\alpha$ line profile subsequently.
The two-component model has five free parameters: the redshift tied between narrow and broad H$\alpha$ component, and line widths and line luminosities of both narrow+broad components.
We use the python implementation of MCMC \citep[\texttt{emcee};][]{emcee} to estimate the posteriors of the five free parameters, taking the 50th percentile as the best estimation and 16th- to 84th-percentile range as the uncertainties.
The UV luminosity $\nu L_{\nu,1500}$ is directly measured from the spectrum at $\lambda_{\rm rf}=1500$ \AA.

We also apply a dust attenuation correction, so that we can examine the intrinsic energy budget between rest UV light and ionizing radiation and explore the ionizing sources in LRDs and LBDs.
Although their UV colors are relatively blue by definition, they are not as blue as dust-free ($\langle\beta_{\rm UV}\rangle=-1.3$) and suggest a modest dust attenuation.
We therefore use the rest UV color $\beta_{\rm UV}$ as the probe of dust attenuation, assuming the Calzetti extinction law \citep[$A_{1600}=2.31\beta_{\rm UV} + 4.85$;][]{Calzetti2000ApJ}.
The dust attenuation estimated from the UV slope is typically small for LRDs ($\langle A_V\rangle=0.7$ mag), which does not violate the non-detection of hot/cold dust emissions of LRDs in JWST/MIRI or ALMA \citep{Casey_2025,K.Chen_2025}.

\section{Results}\label{sec:results}
\subsection{Energy balance between UV and broad/narrow lines in LRDs and LBDs}\label{subsec:UVHa}

\begin{figure*}[t]
\centering
\includegraphics[width=0.8\textwidth]{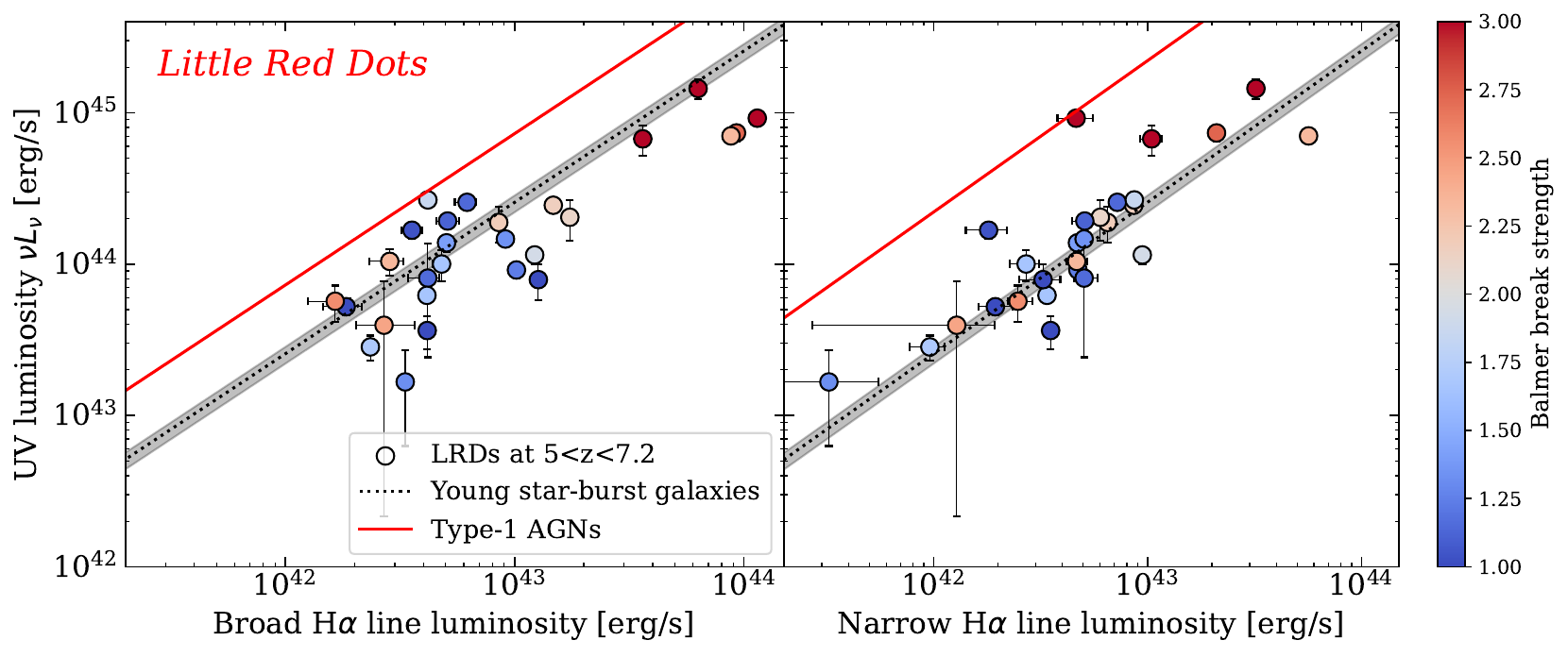}
\includegraphics[width=0.8\textwidth]{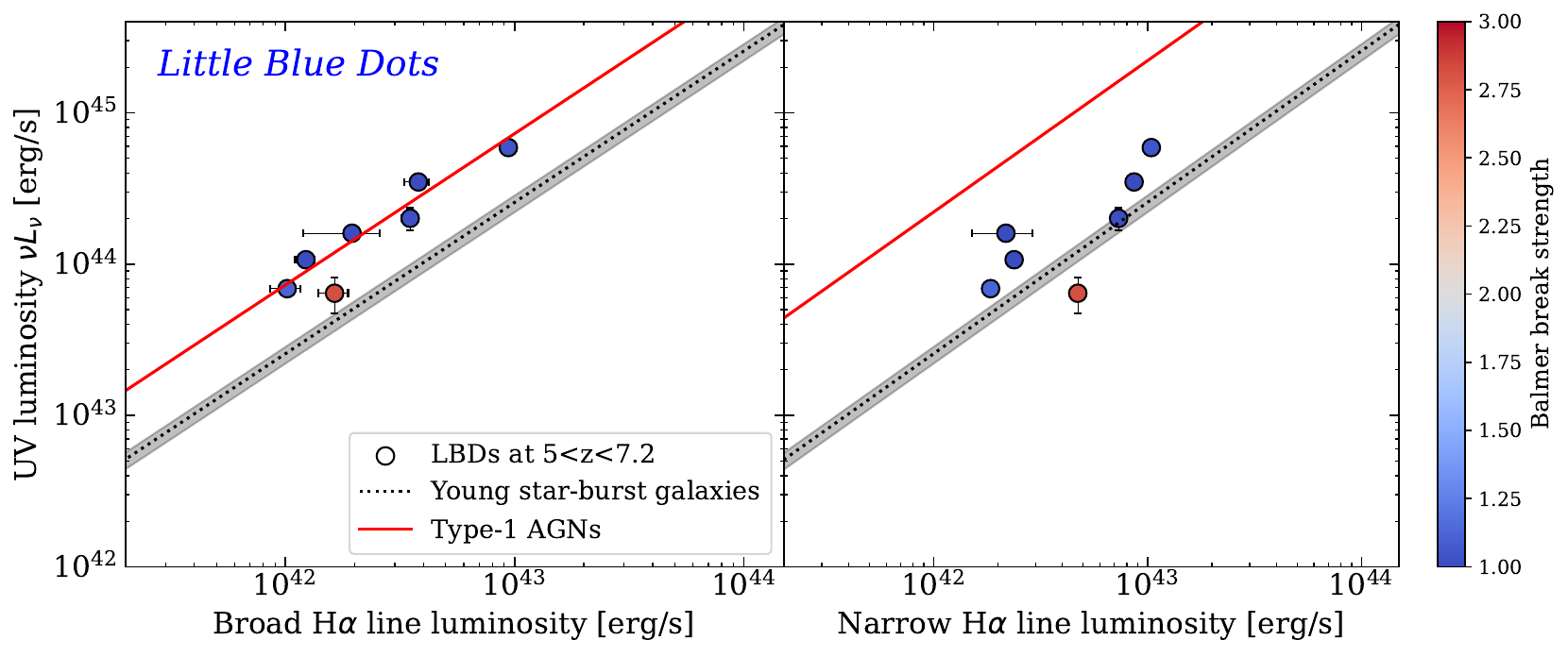}
\caption{Rest UV versus H$\alpha$ line luminosity in LRDs (top) and LBDs (bottom). Broad and narrow components are displayed on the left and right, respectively. Plots are color-coded by the Balmer break strength, defined as $f_{\nu,4230}/f_{\nu,3560}$. As a reference, the UV-to-H$\alpha$ luminosity ratios for young starburst galaxies (black dotted line) and typical Type 1 AGNs (red solid line) are also shown. The gray shaded band shows the range of the luminosity ratio when the stellar age and gas density is varied for young star-forming galaxies.
}
\vspace{5mm}
\label{fig:UV_Ha}
\end{figure*}

Figure~\ref{fig:UV_Ha} presents the UV and H$\alpha$ luminosities of LRDs (top) and LBDs (bottom) with the broad and narrow H$\alpha$ components separately in the left and right panels.
The UV and broad/narrow H$\alpha$ luminosities are measured from the NIRSpec/MSA spectra, so the energies from the same physical region are compared.
For reference, we overlay the UV-to-H$\alpha$ luminosity ratios for young starburst galaxies (black dotted) and for local AGNs (solid red).
The UV-to-H$\alpha$ luminosity ratio for young starburst galaxies is computed based on \texttt{BPASS} stellar population models \citep{Eldridge2017PASA} combined with photoionization calculations performed with \texttt{Cloudy} \citep[v23.01;][]{Chatzikos2023RMxAA}. 
Specifically, we compute the $\nu L_{\nu,1500}/L_{\textrm{H}\alpha}$ ratios by varying the electron density and stellar age over the ranges of $n_{\rm e}=10^{2.5-4.5}\ {\rm cm^{-3}}$ and $t_{\rm age}=10^{6-7}$ yr. 
For AGNs, the UV-to-H$\alpha$ luminosity ratios are measured directly from the compiled Type 1 AGN spectrum \citep{VandenBerk2001AJ}.
We decompose the H$\alpha$ line profile into narrow and broad components by fitting two Gaussian profiles, and derive the corresponding ratios for each component separately.
Table~\ref{tab:Lum_ratios} collates observations and predictions.

In LRDs, the UV continuum luminosity is correlated with the H$\alpha$ line luminosity for both the narrow and broad components (top panels of Figure~\ref{fig:UV_Ha}).
The correlation is particularly tight for the narrow-line component (top right), 
and the UV-to-H$\alpha$ luminosity ratio is remarkably consistent with that expected for young starburst galaxies.
The Pearson correlation coefficient ($r_{\rm p}$) and $p$-value are $r_{\rm p}=0.67$ and $p=1.2\times10^{-4}$, and the correlation is statistically significant.
The median of UV-to-H$\alpha$ luminosity ratio is $29\pm7$ for LRD narrow lines, which is consistent with young star-burst galaxies ($26^{+3}_{-4}$) but considerably different from normal AGN NLRs ($\sim220$).
For broad H$\alpha$ components in LRDs (top left), although the scatter is slightly larger, the UV and broad H$\alpha$ luminosity are also correlated and follow the similar trend.
The correlation coefficient and $p$-value are $r_{\rm p}=0.83$ and $p=6.5\times10^{-8}$.
The median luminosity ratio is $17\pm8$, which is again consistent with young star-burst galaxies rather than normal AGN BLRs ($\sim70$).

\textit{LBDs} also show clear correlations between the UV continuum and H$\alpha$ luminosities, 
though, the UV-to-H$\alpha$ luminosity ratios in LBDs differ systematically from those in LRDs, 
and the nature of the difference depends on whether the narrow or broad H$\alpha$ component is considered.
For the narrow H$\alpha$ component, the UV-to-H$\alpha$ ratios in LBDs are broadly consistent with those expected for young starburst galaxies, but with larger scatter than observed in LRDs ($\langle L_{\rm UV}/L_{{\rm H}\alpha}\rangle=40\pm12$).
Moreover, the relation is slightly shifted closer to that expected for AGN narrow-line emission, suggesting that AGN-powered emission makes some contribution to the narrow H$\alpha$ lines in LBDs.
By contrast, the broad H$\alpha$ component in LBDs follows a fundamentally different relation.
Its UV-to-H$\alpha$ luminosity ratios ($\langle L_{\rm UV}/L_{{\rm H}\alpha}\rangle=68\pm14$) are incompatible with those of young starburst galaxies ($\sim26$) and instead closely match the relation observed for broad lines in local Type 1 AGNs ($\sim70$).

The fact that the fundamental difference between LRDs and LBDs emerges exclusively in the broad H$\alpha$ component strongly suggests a connection between the engine of BLRs and the presence of optical red continua characterizing V-shaped spectra in LRDs.
The $p$-value from the Kolmogorov-Smirnov test comparing the UV-to-H$\alpha$ broad line luminosity ratios in LRDs versus in LBDs is $1.8\times10^{-4}$, and thus the energy balance between UV and broad H$\alpha$ lines is statistically different depending on the presence of V-shaped continuum.
Note that this statistical difference holds regardless of the dust attenuation correction, and the $p$-value when comparing the luminosity ratios among the two populations before dust attenuation correction is $1.3\times10^{-4}$.
As mentioned above, moreover, the UV-to-H$\alpha$ luminosity ratios are typically similar to those in young star-burst galaxies for sources with V-shaped continuum (LRDs) while their ratios become consistent with BLRs in local Type 1 AGNs for sources without V-shaped continuum (LBDs).
If this red continua originate from a dense gaseous envelope around a growing BH, 
the results imply that the ionizing source of the broad-line clouds transitions from stellar radiation in LRDs to accretion-disk radiation in LBDs, marking a qualitative change in the nature of the central power source (Table~\ref{tab:Lum_ratios}).

\begin{deluxetable}{ccc}
    \label{tab:Lum_ratios}
    \tablecaption{UV-to-H$\alpha$ luminosity ratios
  	}
    \tablewidth{\textwidth}
    \tablehead{
    \colhead{Sample} & \colhead{Broad component} & \colhead{Narrow component}
    }
    \startdata
    LRDs\tablenotemark{$\star$} & $17\pm8$ & $29\pm7$ \\
    LBDs\tablenotemark{$\star$} & $68\pm14$ & $37\pm10$\\
    \hline
    Type 1 AGNs\tablenotemark{$\dagger$} & $\sim70$ & $\sim220$\\
    Young SFGs\tablenotemark{$\dagger\dagger$} & \multicolumn{2}{c}{$26^{+3}_{-4}$} \\
    \enddata
    \tablenotetext{}{$\star$ Median values among the sample are quoted. The standard deviations are quoted as the uncertainty.}
    \tablenotetext{}{$\dagger$ Measured from the compiled spectrum of local Type 1 AGNs by \citet{VandenBerk2001AJ}.}
    \tablenotetext{}{$\dagger\dagger$ Estimated from BPASS+Cloudy photoionization modeling. The possible range of this ratio under different assumptions on nebular/stellar conditions are quoted as the uncertainty.}
\end{deluxetable}



\subsection{Lyman-alpha emission in LRDs and LBDs}\label{subsec:LyA}

We also examine the Ly$\alpha$ emission lines in the sample broad line emitters.
Since Ly$\alpha$ is a resonance line, Ly$\alpha$ photons are unlikely to escape through a dense gas envelope. Therefore, the Ly$\alpha$ occurrence rate and the line strength provide a useful diagnostic of the origin of (narrow) hydrogen recombination lines.
We use the best-fit power-law for the UV continuum (obtained in Section~\ref{sec:data}) as the continuum baseline around the Ly$\alpha$ emission, and we model the Ly$\alpha$ line with a single Gaussian on top of the continuum.
The IGM attenuation is applied on the continuum following the prescription by \citet{Inoue2014MNRAS} and \citet{Asada2025ApJ}, so that the potential rounded shape of the Lyman break can also be parameterized.
We then identify Ly$\alpha$ emitters (LAEs) among the sample broad H$\alpha$ emitters by searching for Ly$\alpha$ lines with S/N$>3$.
Sources with Ly$\alpha$ detections are marked with squares in Figure~\ref{fig:sample} (a), and the two examples shown in panel (b) and (c) are both LAEs.

The Ly$\alpha$ occurrence rates are $37^{+16}_{-11}\ \%$ (10/27) and $14^{+33}_{-12}\ \%$ (1/7) among LRDs and LBDs.
These rates are statistically consistent with each other, or LRDs may have even higher occurrence rate than LBDs, indicating that the presence of a dense gas envelope does not suppress the observability of Ly$\alpha$ photons in these two classes.
Moreover, the occurrence rate of $\sim20-40\%$ is comparable to that observed in normal star-forming galaxies (SFGs) at similar redshifts and UV magnitudes ($M_{\rm UV}\simeq -19$ mag; e.g., \citealt{Stark2011ApJ,DeBarros2017AAP,Kusakabe2020AAP,Goovaerts2023AAP,Napolitano2025arXiv}).
The similar Ly$\alpha$ occurrence rates in LRDs, LBDs, and normal SFGs further suggest that the narrow lines originate outside the dense gas envelope, most plausibly from H~{\sc ii} regions within the host galaxies.

\begin{figure}[t]
\centering
\includegraphics[width=0.96\columnwidth]{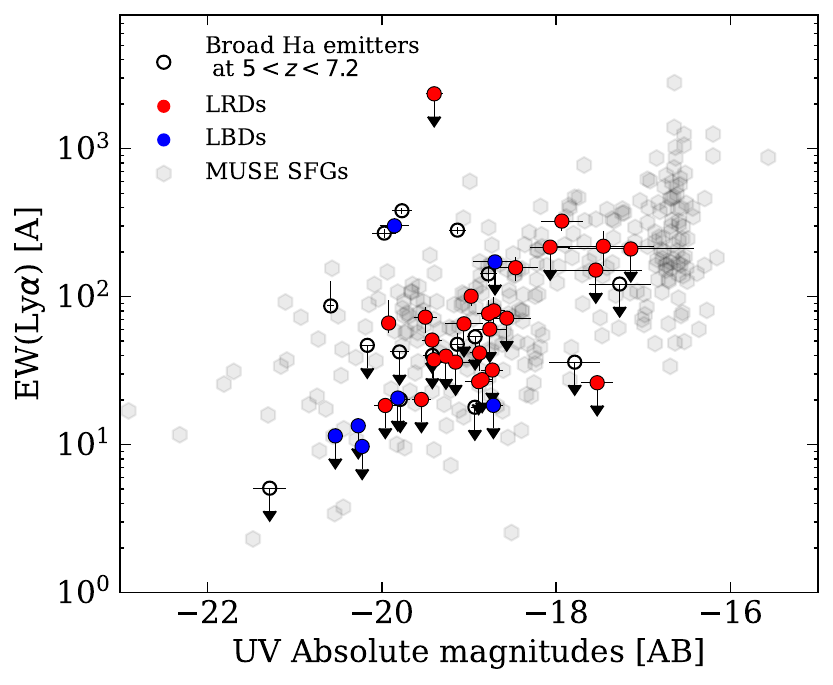}
\caption{Ly$\alpha$ line equivalent widths of the sample broad H$\alpha$ emitters at $z=5-7.2$, plotted against the rest UV absolute magnitude. They distribute at a very similar locus as normal SFGs at the same redshift range \citep[][]{Kerutt2022AAP}, and there is no significant difference between LRDs and LBDs.
}
\label{fig:LyaEW}
\end{figure}

Figure~\ref{fig:LyaEW} presents the Ly$\alpha$ equivalent width of the sample, plotted against their rest UV absolute magnitudes.
No systematic difference can be seen between LRDs and LBDs, and they occupy very similar locus in the EW(Ly$\alpha$)-$M_{\rm UV}$ plot as normal SFGs at the same radshift range identified with VLT/MUSE \citep{Kerutt2022AAP}.
Considering AGN-powered Ly$\alpha$ emission can be more luminous than normal SFGs \citep[e.g.,][]{Sobral2018MNRAS}, there seems no strong evidence supporting that the (narrow) Ly$\alpha$ lines in LRDs or LBDs are mainly powered by AGN accretion.
We should caution that, however, our search for the Ly$\alpha$ emission with NIRSpec/Prism spectra could be incomplete particularly for low EW(Ly$\alpha$) sources due to the low spectral resolution of the Prism at these short wavelengths. It is also difficult to compare the intrinsic emission line profiles between Ly$\alpha$ and H$\alpha$ narrow component due to the low spectral resolution, which could otherwise be informative to probe the origin of the Ly$\alpha$ line engine and escape mechanisms \citep{Baek2013MNRAS,Yang2014ApJ}.
Further large spectroscopic campaigns of LRDs and LBDs at rest UV wavelengths with higher spectral resolution are needed to improve the statistics.




\section{Discussion}

\subsection{Gradual transition between LRDs and LBDs}\label{subsec:continuity}
The statistical analyses of LRDs and LBDs in the previous section demonstrate the fundamental difference between LRDs and LBDs, particularly in the energy balance of the broad H$\alpha$ line vs the rest UV continuum luminosity.
It is thus essential to examine whether there is an evolutionary link between LRDs and LBDs or they are distinct unrelated population.
To this end, we select the intermediate population between LRDs and LBDs in the $\beta_{\rm opt}$-$\beta_{\rm UV}$ diagram (Figure \ref{fig:sample}a) and measured their $L_{\rm UV}/L_{ {\rm H}\alpha, {\rm broad}}$ ratios.
Namely, we define the intermediate population as:
\begin{enumerate}
    \item $\beta_{\rm opt}<0$,
    \item $\beta_{\rm opt} - \beta_{\rm UV} > 0.5$,
\end{enumerate}
and perform the same two-component Gaussian fitting as we do for LRDs/LBDs.
A total of 12 intermediate-color population is selected.

\begin{figure}[t]
\centering
\includegraphics[width=0.96\columnwidth]{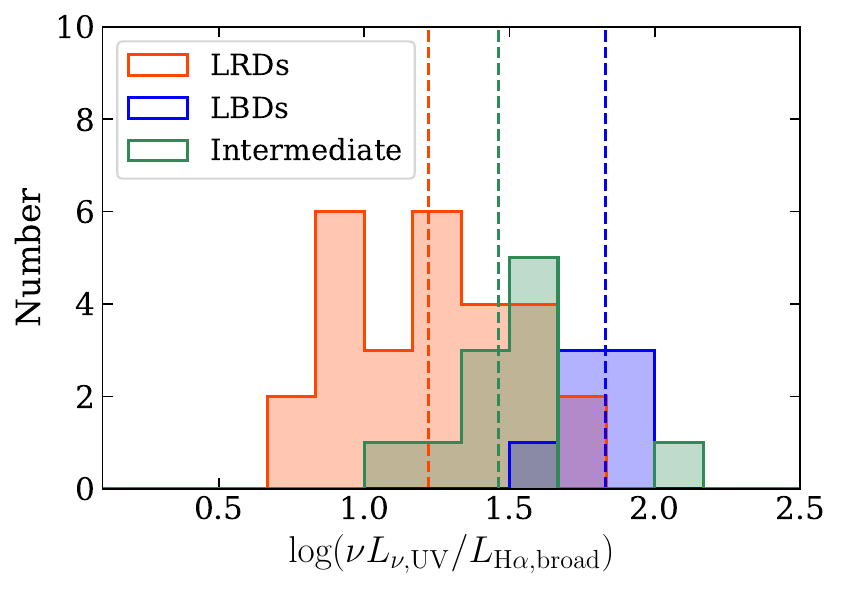}
\caption{Histograms of the UV-to-H$\alpha$ luminosity ratios for the broad-line component in LRDs (red), LBDs (blue), and the intermediate-color population (green). The vertical dashed line denotes the median of each population.
}
\label{fig:UVHa_hist}
\end{figure}

The intermediate-color population shows an intermediate property between LRDs and LBDs.
The UV-to-H$\alpha$ ratios for the broad component in this population are typically $\langle L_{\rm UV}/L_{{\rm H}\alpha}\rangle=29\pm7$, which is larger than LRDs but smaller than LBDs.
Figure \ref{fig:UVHa_hist} presents the histogram of broad line UV-to-H$\alpha$ ratios of the three populations, and it demonstrates the gradual transition between LRDs and LBDs, connected via the intermediate-color population.
This obviously suggests that the LRD is not a distinct population, but rather it should be interpreted as an evolutionary stage linked with the normal, bule AGNs (LBDs).
Although it is not fully clear which direction they evolve (LRDs to LBDs or LBDs to LRDs), there must be a link between the UV-to-optical color and broad Balmer line emissivity, and LRDs are gradually realized with respect to the normal blue AGN population.

\subsection{Constraint on LRD models from spectroscopic observations}\label{subsec:validate_models}

In this section, we discuss the constraint on the models for LRDs from the observational results presented in the previous section.
The observed correlations between the UV continuum and H$\alpha$ lines, together with their luminosity ratios 
and Ly$\alpha$ occurrence rates, provide key diagnostics of the dominant ionizing sources.
We examine whether various envelope models (or ``BH*'' models) in the literature can consistently account for these observational constraints and assess their relative plausibility.
We mainly focus on the envelope models proposed in literature (see Figure~\ref{fig:models}), but also briefly mention the possibility of other models than envelope models.

\vspace{4pt}
$\bullet$ {\it Blackbody envelope model}\\
In this model (Figure~\ref{fig:models}a), the gaseous envelope surrounding the central BH is completely opaque and produces a blackbody radiation spectrum
\citep[e.g.,][]{Kido_2025,Begelman_2025,Santarelli2026ApJ}.
As a result, the rest-frame UV continuum as well as both narrow and broad hydrogen lines in LRDs 
are predominantly powered by young massive stars residing in the compact star cluster ($\lesssim 10-100~{\rm pc}$) together with the central LRD,
rather than by the AGN accretion disk 
(Figure~\ref{fig:models}a; \citealt{Inayoshi2025arXiv}).
Since the UV continuum and ionizing radiation responsible for H$\alpha$ lines share a common stellar origin, this model 
naturally explains the observed tight correlations between UV and H$\alpha$ (narrow and broad) line luminosity and their luminosity ratios. 
The (narrow) Ly$\alpha$ lines also arise from H~{\sc ii} regions associated with the compact stellar cluster in this model, which can account for the observed occurrence rate and Ly$\alpha$ strengths that are comparable to normal SFGs at the same redshift range.

We also note that the stellar origin for the UV continuum can account for observations of LRDs spatially resolved in the rest-frame UV,
comprising $\sim 1/3$ of the samples reported in the literature \citep[e.g.,][]{Rinaldi2025ApJ,Kokorev2025arXiv,Baggen2026ApJ}.
The remaining LRDs are spatially unresolved, including gravitationally lensed sources with tighter size limits ($<100~{\rm pc}$)
yet show V-shaped SEDs \citep[e.g.,][]{Furtak2024Natur,C.Chen_2025,Yanagisawa2026arXiv}, which suggests that the UV emission mainly originates from young compact stellar clusters in the nuclear region, 
as described in this model (Figure~\ref{fig:models}a). 
Additionally, this picture is consistent with a recent work by \citet{Sun2026arXiv} that shows that the UV luminosity in LRDs can be predominantly attributed to the SFG component, by decomposing the LRD spectra into SFG and central engine component.
Overall, this model provides a self-consistent explanation for the observational results and is thus the most favored 
among the envelope models considered here.

\vspace{4pt}
$\bullet$ {\it Cocoon envelope model}\\
In this model (Figure~\ref{fig:models}b), the UV continuum and narrow emission lines originate in the host galaxy, 
while the broad-line emission is powered by the AGN accretion disk deeply embedded within a dense cocoon 
(Figure~\ref{fig:models}b; \citealt{Naidu_2025,deGraaff_2025b}).
This scenario can explain the tight correlation of the UV and the narrow H$\alpha$ luminosity in LRDs. 
The stellar origin of the UV light in this model can also account for the morphology similar to the blackbody envelope model case.
However, this model has a difficulty in replicating the observed correlation between the UV continuum and the broad H$\alpha$ luminosity, and a very fine-tuned geometry and non-trivial coincidence are needed to explain the observations.
The observed UV-to-broad H$\alpha$ correlation might be attributed to coevolution between the host galaxy and the central BH.
In this case, though, a certain degree of fine tuning is invoked such that the rest UV and broad H$\alpha$ luminosities are correlated and their luminosity ratios coincidentally resemble those of young starburst galaxies.
Moreover, the cocoon model requires a finely tuned density, clumpiness, and geometric structure of the surrounding gas.
The cocoon must be dense enough to absorb the UV continuum from the accretion disk, yet allow Balmer emission lines to escape from the inner ionized region without being full thermalization.
Therefore, while the cocoon model may not be completely ruled out, it relies on several non-trivial assumptions to explain the full set of observations, and it should be less likely than other scenarios.

\vspace{4pt}
$\bullet$ {\it Non-spherical envelope model}\\
In the non-spherical gas configuration (Figure~\ref{fig:models}c), the UV continuum and both the narrow and broad hydrogen Balmer lines are 
predominantly powered by the AGN accretion disk \citep[e.g.,][]{Lin_2025c}\footnote{The non-spherical envelope model is defined as the case where both the UV continuum and the reprocessed emission lines powered by the AGN can escape through low-density, polar directions. Although \citet{Lin_2025c} adopt a non-spherical gas geometry and provide a schematic illustration, we here note that they do not explicitly specify the origin of the UV continuum, treating it as an open question.}.
A common AGN power source naturally explains the correlation between UV and H$\alpha$ luminosities.
However, this model cannot account for the systematic differences in UV-to-H$\alpha$ luminosity ratios between LRDs and LBDs without tuning additional configuration, as it does not naturally predict a significant change in the energy budget associated with the presence or absence of the envelope.
Furthermore, the similarity of the Ly$\alpha$ occurrence rates in LRDs, LBDs, and normal SFGs are not favorable to this scenario.

A possible explanation for the observed enhancement of H$\alpha$-to-UV luminosity ratio under this hypothesis is a higher gas covering fraction in the nuclear region during the LRD phase.
Near the surface of the dense gaseous envelope (on spatial scales comparable to the BLR sizes inferred from reverberation mapping in nearby AGNs), BLR clouds occupy a large fraction of the solid angle of the radiation-escape funnel,
thereby reprocessing ionizing photons into line emission more efficiently \citep[e.g.,][]{Maiolino_2025_Xray,Yan_2025}.
Therefore, the non-spherical envelope model could explain the different UV-to-H$\alpha$ luminosity ratios between LRDs and LBDs if the BLRs during the LRD phase are assumed to have higher covering fractions and larger gas column densities than normal unobscured AGNs (LBDs).
However, this assumption introduces a larger scatter in the UV-H$\alpha$ relation and requires LBDs to have lower covering fractions.
It also remains physically unclear why the offset between LRDs and LBDs appears at a factor of $\sim 4$.

In any case, a non-spherical envelope phase may still play an important role during the evolutionary transition from LRDs to normal AGNs \citep[see e.g.,][for examples of transition-phase LRDs]{Merida2025arXiv}. The continuous distribution observed in the color–color diagram (Figure~\ref{fig:sample}a) from LRDs to LBDs suggests that LRDs can gradually evolve into normal Type 1 AGNs \citep[see also, e.g.,][for a similar discussion on the evolutionary stages of LRDs]{Barro2025arXiv}. Such a transition could naturally proceed through a non-spherical envelope phase, providing a physical link to the unified model of AGNs \citep[e.g.,][]{Netzer2015ARAA}.

\vspace{4pt}
$\bullet$ {\it Non-envelope models}\\
Although recent theoretical and observational works often focus on envelope models, they were originally proposed to explain a subset of LRDs with extreme properties such as deep Balmer breaks, Balmer absorption lines, exponential wing line profiles, or blackbody-like continuum shape from rest optical to infrared \citep[e.g.,][]{Naidu_2025,Rusakov2025arXiv,Lin_2025c}. 
It is thus possible that the dense gas envelope is needed only for these extreme LRDs and different physics may lie behind the overall normal population of LRDs.
However, our LRD sample selected in this paper includes some of these extreme LRDs, which requires a dense gas envelope to explain its properties -- three shows a Balmer break deeper than a factor of 3, and a roughly half of the sample LRDs indeed show considerably flattening in the rest-optical slope at $>5500$ \AA\ (i.e., a blackbody-like continuum shape).
There is no clear evidence suggesting these LRDs are distinctive/disconnected from other normal LRDs. 
The color distribution in the $\beta_{\rm opt}$-$\beta_{\rm UV}$ diagram and the gradual transition in the UV-to-H$\alpha$ luminosity ratios (Section \ref{subsec:continuity}) suggest that they just occupy the high-end tail of a continuous distribution rather than disconnected from the rest.
This should indicate that the envelope models are more plausible for the entire LRD population than others, as they do not require additional non-trivial fine-tuning.

We finally note that our key results presented in Section \ref{sec:results} do not depend on the envelope model assumption.
The sample selection of LRDs and LBDs are purely based on the rest UV and optical colors among broad H$\alpha$ line emitters.
The correlation between UV continuum and narrow/broad H$\alpha$ lines and the Ly$\alpha$ emissivity are not affected by the assumption of the origin of the red rest optical continuum in LRDs.
Therefore, the main results in this paper are important constraints on any successful LRD model.
The systematic offset in the UV-to-H$\alpha$ luminosity ratio between LRDs and LBDs needs to be accounted even when other models are considered.

\subsection{Future tests of envelope models}
Although the blackbody envelope model is currently the most preferred, other types of envelope models are not fully ruled out, and these models can be further validated with future observations.
One of the promising approaches is to explore the time variability of LRDs \citep[e.g.,][]{Furtak_2025,Kokubo_Harikane_2025,Ji_2025}.
Since emissions powered by AGN accretion can have time variability while stellar-origin emissions are not or weakly variable, we can distinguish which emission components in LRDs originate from AGNs or stars with their variability.
For example, the blackbody envelope model predicts time variability only in the rest optical continuum but not in the emission line luminosities, while the cocoon model predicts both in the rest optical continuum and the broad line luminosity.
Recently, \citet{Zhang2025arXiv} utilized the time delay effect between multiple images of one single LRD due to the strong gravitational lensing and reported long-term ($\sim 100~{\rm yr}$ in the observer frame) variability in the photometric colors, yet it is not clear whether the broad line component also shows the variability.
Deep spectroscopic observations of multiply-lensed LRDs will thus be an important diagnostic to validate the envelope models.

Another key observations would be to probe the structures of BLRs in LRDs.
Normal AGNs in the local universe are known to have a stratified ionized structure in BLRs, where higher ionization potential broad lines reside in the inner radii \citep[e.g.,][]{Peterson2006LNP,Gaskell2009NewAR,Kollatschny2013AAP}.
A few previous studies on LRDs found that broad low ionization-potential lines have lower FWHMs than broad Balmer emission lines (e.g., O~{\sc i}~$\lambda$8448 or Fe~{\sc ii}; \citealt{Kokorev2025arXiv,Lin_2025c,Tripodi2025ApJ}), suggesting that BLRs in LRDs could also have similar stratified structures.
This can be naturally explained by the cocoon envelope model or the non-spherical envelope model, while an extra explanation is needed in the blackbody envelope model.
However, the stratified BLR structures in LRDs have been observed only in lines with ionization potentials lower than Balmer lines that can also be emitted by Ly$\alpha$/Ly$\beta$ fluorescence process, which makes the detailed ionization structures in LRD BLRs remain unclear.
In particular, finding high ionization-potential lines such as C~{\sc iv}$~\lambda$1550 with broader line widths than Balmer lines in LRDs would place a severe constraint against the blackbody envelope model.

\subsection{Duty cycles of LRDs}

Our analysis indicates that the UV continuum observed in LRD spectra originate from the stellar emissions. 
If this is the case, the UV luminosity function (LF) of LRDs directly reflects that of their stellar component.
Previous studies with photometrically selected LRDs \citep{Kokorev_2024a,Kocevski2025ApJ} showed that the UV LF of LRDs at $z=4.5-6.5$ is approximately two orders of magnitude lower in normalization than that of Lyman-break galaxies 
at the same redshifts, while the LF shapes are remarkably similar across a wide magnitude range \citep[from $M_{\rm UV}\simeq -18$ to $-21$ mag; e.g.,][]{Bouwens2021AJ,Harikane2022ApJS}.

The similar LF shapes but different normalizations suggest that only $\sim 1\%$ of rest-UV-selected star-forming galaxies at 
$z\sim 5-6$ appear as LRD host galaxies.
This fraction can be interpreted as the duty cycle of LRDs, yielding $f_{\rm duty}\sim1\%$ at these redshifts.
Such a duty cycle corresponds to a typical LRD lifetime of $\sim 1\%$ of the Hubble time, or $\sim 12~[(1+z)/6]^{-3/2}$ Myr.
The estimate agrees well with an independent constraint on $f_{\rm duty}$ based on the characteristic timescale of stellar feedback, 
which is expected to quench gas supply to the nuclei and thus disperse the thick envelope structure \citep{Inayoshi2025arXiv}.
We note that this duty cycle estimation implicitly assumes that all (UV-selected) galaxies randomly experience the short-lived LRD-hosting phase.

An alternative interpretation of the $\sim2$ dex lower normalization is a consequence of the small fraction of galaxies that can be an LRD-host for the entire life.
These two possible interpretations are contrasting in the sense that LRDs can be hosted in ubiquitous galaxies (dusty cycle scenario) or only in anomaly galaxies (LRD-host fraction scenario).
In reality, LRDs should have their lifetime considering the stark decline in the number density in the lower redshift \citep[e.g.,][]{Inayoshi2025ApJ,Ma2026ApJ} and the combination of these two scenarios could be the case. Nonetheless, future characterization of the LRD-host galaxies would be necessary to conclude which of the two interpretations is more likely.

\begin{figure}[t]
\centering
\includegraphics[width=0.96\columnwidth]{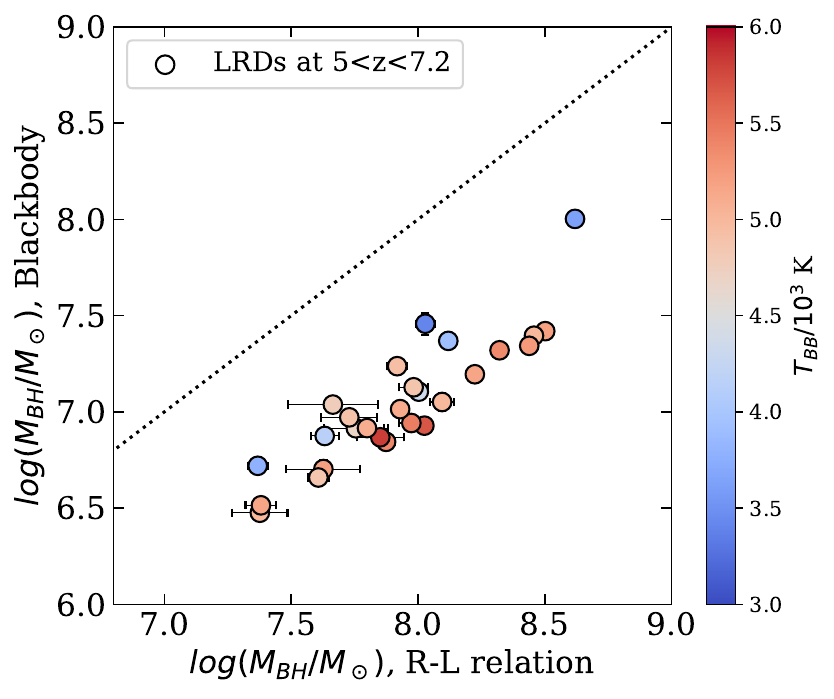}
\caption{The effect of different size-luminosity relation on the BH mass estimation.
The figure compares the BH mass estimations of the sample LRDs, assuming the typical $R_{\rm BLR}$-$L$ relation calibrated for AGN accretion disk ($x$-axis) and assuming that BLRs are just at the photospheric radius of the blackbody envelope ($y$-axis).
Plots are color-coded by the blackbody temperature ($T_{\rm BB}$).
It demonstrates that using the typical $R_{\rm BLR}$-$L$ relation can overestimate the BH mass by $\sim1$ dex due to the different $R_{\rm BLR}$ estimations.
}
\label{fig:BH_mass}
\end{figure}

\subsection{BH mass estimator for LRDs}

In the blackbody envelope model (Figure~\ref{fig:models}a), even the broad Balmer emission lines are not primarily powered by the AGN accretion disk, 
but instead by young massive stars in the compact stellar cluster.
As a result, no direct emission from the accretion disk is observable, and single-epoch virial BH mass estimates based on 
the $R_{\rm BLR}$-$L$ relation calibrated by reverberation mapping of nearby AGNs become inappropriate \citep[e.g.,][]{Greene2005ApJ,Bentz2013ApJ}.
Instead, the broad-line clouds illuminated by stellar radiation reside near the surface of the gaseous envelope and are gravitationally bounded by the central BH.
Thus, the characteristic radius of the BLR can be approximated by the photospheric radius of the envelope $R_{\rm ph}$, allowing the BH mass to be inferred from  $R_{\rm ph}$ and the line width of the broad H$\alpha$ emission line.

Figure~\ref{fig:BH_mass} compares the BH mass measurements of our LRD sample derived from the standard $R_{\rm BLR}$-$L$ relation and those inferred from 
the blackbody photospheric radius.
In the former case, we estimate BH masses following the prescription of \citet{Reines2013ApJ}, which uses the broad H$\alpha$ luminosity 
as a proxy for $R_{\rm BLR}$ under the assumption of the standard AGN $R_{\rm BLR}$–$L$ relation (the $x$-axis of Figure~\ref{fig:BH_mass}). The resulting $R_{\rm BLR}$ in the sample LRDs are typicall $\sim6000$ au ($\sim40$ lightdays).
In the latter case, we measure the blackbody temperature ($T_{\rm BB}$) and the total envelope luminosity ($L_{\rm ph}$) by fitting blackbody spectra to 
the Prism spectra of the sample LRDs, and estimate $R_{\rm ph}$ using the Stefan-Boltzmann law.
Assuming that the broad-line clouds are virialized, we then compute the corresponding BH mass with $R_{\rm ph}$ 
(the $y$-axis of Figure~\ref{fig:BH_mass}). 
Plots are color-coded by the best-fit blackbody temperature. It shows that the BB temperatures are tightly clustered around $5000$ K independently of the BH mass, which is in line with the prediction by the blackbody envelope model \citep[e.g.,][]{Kido_2025,deGraaff_2026,Umeda_2025}.
The envelope size estimated from the photospheric radius is typically $\sim900$ au ($\sim5$ lightdays).

The comparison demonstrates that BH masses inferred under the blackbody envelope model are systematically lowered by one order of magnitude
than those obtained using the conventional single-epoch method.
This reduction relieves the ``overmassive" nature of BHs in LRDs relative to the local BH-host mass correlations \citep[e.g.,][]{Furtak2024Natur,Maiolino2024AAP,Kocevski2025ApJ}.
We highlight that this decrease rises solely from the fundamentally different BLR size estimates, rather than from additional line broadening 
due to electron or resonant scattering \citep[e.g.,][]{Rusakov2025arXiv}.
It is thus possible that the BH mass may be even smaller by a further $\sim1$ dex, if the additional line broadening due to scattering takes place in the blackbody envelope model.

\section{Summary}\label{sec:summary}

In this paper, we examine the origins of rest UV light and narrow/broad hydrogen Balmer lines in LRDs.
Recent observational and theoretical works proposed models involving a dense gaseous envelope enshrouding the AGN to explain many of the puzzling features of LRDs, including the red rest-optical continuum, large Balmer decrements, deep Balmer breaks, high equivalent width H$\alpha$ broad emission lines, and lack of X-ray and hot/cold dust emissions.
However, the origins of the UV continuum and broad+narrow emission lines are under debate, and several types of envelope models have been proposed, each predicting distinct energy sources for the UV continuum, narrow lines, and broad lines (Figure~\ref{fig:models}). 
We utilize all archived JWST observational data available on the DAWN JWST Archive (DJA) to make a large sample of spectroscopically-confirmed LRDs, and examine the energy balance between the rest UV continuum and broad/narrow H$\alpha$ line luminosity to constrain the engines.
We also make a control sample of blue, unobscured AGNs at the same redshift range (which we call little blue dots, ``{LBDs}") to explore the effect of gaseous envelopes on the UV-to-H$\alpha$ energy balance.
Our key findings are:
\begin{enumerate}
    \item In LRDs, the UV continuum luminosity is correlated with both narrow and broad H$\alpha$ luminosity. The Pearson correlation $p$-values are $p=1.2\times10^{-4}$ and $6.5\times10^{-8}$, respectively, and their correlations are statistically significant.
    Importantly, the UV-to-H$\alpha$ luminosity ratios are remarkably consistent with young starburst galaxies rather than the typical Type 1 AGN spectrum (Figure~\ref{fig:UV_Ha}, Table~\ref{tab:Lum_ratios}, and Section~\ref{subsec:UVHa}).
    \item In LBDs, the UV continuum is also correlated with both narrow and broad H$\alpha$, but the UV-to-H$\alpha$ luminosity ratio is systematically different from that in LRDs.
    The typical UV to broad H$\alpha$ luminosity ratio in LBDs closely matches the relation observed local Type 1 AGNs.
    The $p$-value from the Kolmogorov-Smirnov test comparing the $L_{\rm UV}/L_{{\rm H}\alpha, {\rm broad}}$ ratios among LRDs versus LBDs is $p=1.8\times10^{-4}$, suggesting that the energy balance between UV and broad H$\alpha$ lines is statistically different depending on the presence of red optical continua
    (Figure~\ref{fig:UV_Ha}, Table~\ref{tab:Lum_ratios}, and Section~\ref{subsec:UVHa}).
    \item The Ly$\alpha$ line occurrence rates in LRDs and LBDs are $37^{+16}_{-11}\ \%$ and $14^{+33}_{-12}\ \%$, respectively, and there is no systematic difference depending on the presence of dense gas envelopes invoking the red rest optical continuum. The occurrence rates and the Ly$\alpha$ equivalent widths of LRDs/LBDs are quite similar to what was found in normal star-forming galaxies at the same redshift range (Figure~\ref{fig:LyaEW}, Section~\ref{subsec:LyA}).
    \item Given these observational results, the blackbody envelope model \citep[e.g.,][]{Inayoshi2025arXiv} is the most favored, where the BH is enshrouded by a completely opaque gaseous envelope, and all UV, narrow, and broad lines are predominantly powered by young massive stars in the compact nuclear region. The cocoon envelope model \citep[e.g.,][]{Naidu_2025}, which predicts the AGN-powered broad lines escaping through the dense gas cocoon, is not completely rejected, although it requires several non-trivial and fine-tuned assumptions to explain the full set of observations (Section~\ref{subsec:validate_models}).
    \item Given the continuous distribution of LRDs and LBDs in the $\beta_{\rm opt}$-$\beta_{UV}$ diagram and the UV-to-H$\alpha$ ratios (Figure \ref{fig:UVHa_hist} and Section \ref{subsec:continuity}), LRDs are not a distinct population, but rather they could gradually evolve to normal blue unobscured AGNs. Non-spherical envelope models \citep[e.g.,][]{Lin_2025c} may play an important role during this transition phase from LRDs to normal AGNs and provide a physical link to the unified model of typical AGNs \citep[e.g., see][]{Netzer2015ARAA}.

\end{enumerate}

A large statistical study holds the key to unveil the mysterious nature of the LRDs.
We demonstrate that the UV continuum, narrow, and broad hydrogen lines in LRDs share a common origin, which is more likely young massive stars rather than AGN accretion.
However, the evolutionary pathways from LRDs to normal Type~1 AGNs remain largely unexplored to date.
A larger spectroscopic sample of broad line emitters, not only LRDs but also LBDs, is crucial to cover the wide area in the $\beta_{\rm opt}$-$\beta_{\rm UV}$ diagram.
This will further enable us to explore the evolutionary paths of LRDs.

\begin{acknowledgments}
We thank Hollis Akins and Steve Finkelstein at the University of Texas at Austin, Changhao Chen at Kavli Institute for Astronomy and Astrophysics, Peking University, and Xiaojing Lin at Tsinghua University for fruitful discussions on this research.
The authors acknowledge the support and hospitality of the Institute for Fundamental Physics of the Universe  (IFPU), whose Focus Week Program "Unraveling Little Red Dots: Linking JWST Discoveries with Simulations to Understand Early Galaxies” organized by R. Tripodi contributed to the development of this work.
This work is based on observations made with the NASA/ESA/CSA JWST.
The data products presented herein were retrieved from the Dawn JWST Archive (DJA). DJA is an initiative of the Cosmic Dawn Center (DAWN), which is funded by the Danish National Research Foundation under grant DNRF140.
This research used the Canadian Advanced Network For Astronomy Research (CANFAR) operated in partnership by the Canadian Astronomy Data Centre and The Digital Research Alliance of Canada with support from the National Research Council of Canada, the Canadian Space Agency, CANARIE and the Canadian Foundation for Innovation.
The Dunlap Institute is funded through an endowment established by the David Dunlap family and the University of Toronto.
Y.A. is supported by the Dunlap Institute.
K.I. acknowledges support from the National Natural Science Foundation of China (12573015, W2532003), 
the Beijing Natural Science Foundation (IS25003), and the China Manned Space Program (CMS-CSST-2025-A09).

\end{acknowledgments}





%
\facilities{JWST (NIRCam and NIRSpec)}

\software{astropy \citep{2013A&A...558A..33A,2018AJ....156..123A,2022ApJ...935..167A},  
          Cloudy \citep{2013RMxAA..49..137F}, 
          msaexp \citep{Brammer2022zndo}
          }


\appendix

\section{List of LRDs and LBDs used in this work}\label{apx:list}
This section lists the LRDs and LBDs analyzed in this work.
Table~\ref{tab:sample_lrds} present the basic properties of all LRDs and LBDs, selected from DJA in this work. The UV luminosity and broad/narrow H$\alpha$ luminosities are corrected for the dust attenuation (see Section \ref{sec:data}).

\begin{sidewaysfigure*}[t]
\centering
\includegraphics[width=0.9\textwidth]{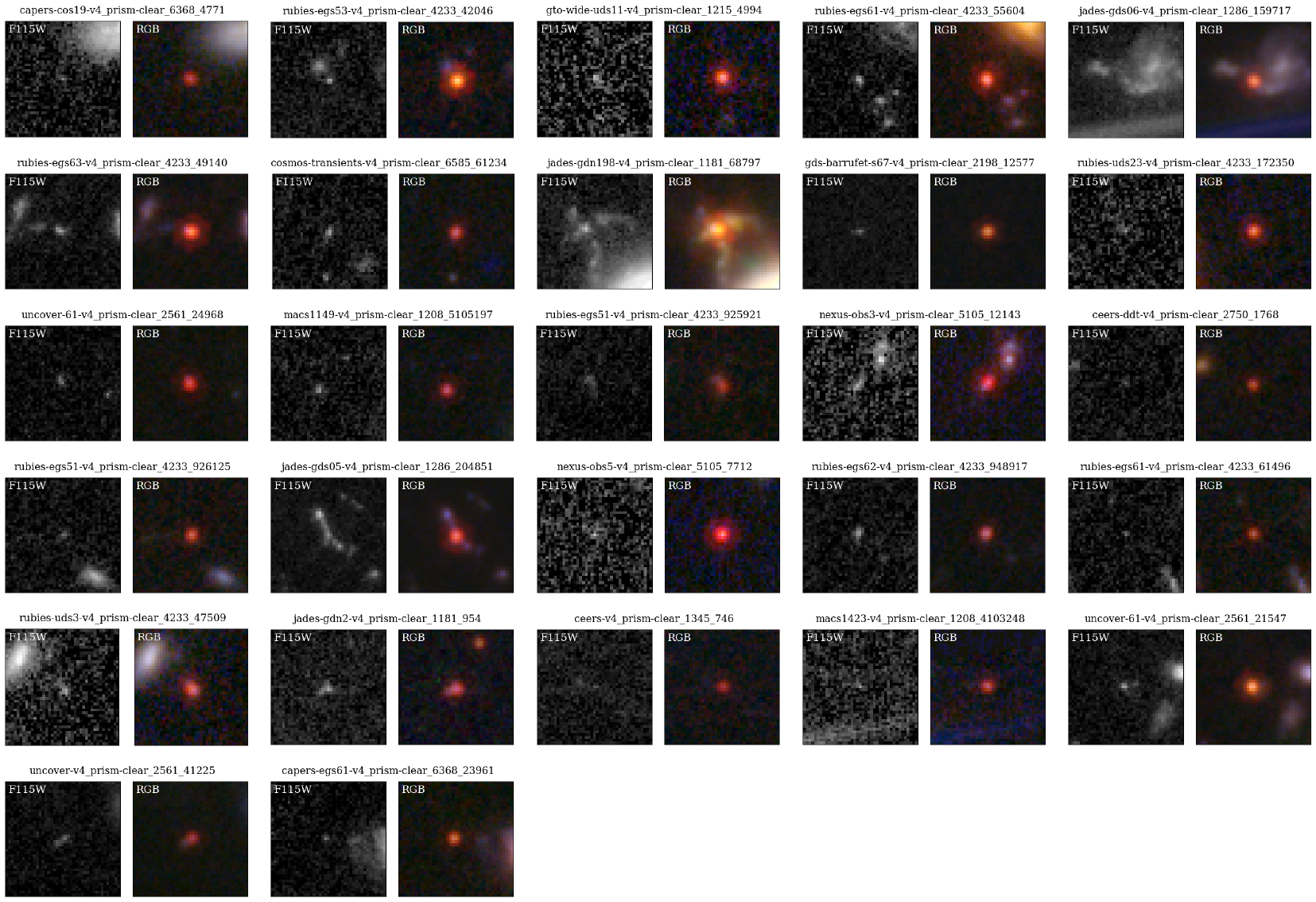}
\caption{NIRCam images of LRDs in this work. For each source, NIRCam/F115W and the composite RGB image are shown.
}
\vspace{5mm}
\label{fig:thumbs_LRDs}
\end{sidewaysfigure*}

\begin{figure*}[t]
\centering
\includegraphics[width=0.8\textwidth]{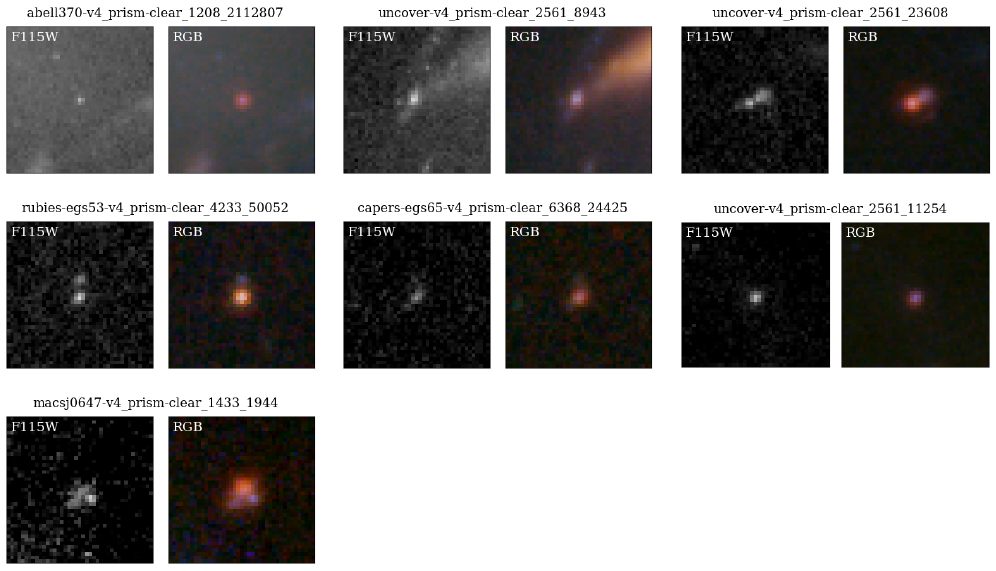}
\caption{Same as Figure \ref{fig:thumbs_LRDs} but for LBDs.
}
\label{fig:thumbs_LBDs}
\end{figure*}

Figure \ref{fig:thumbs_LRDs} and \ref{fig:thumbs_LBDs} shows the NIRCam image cutouts of the LRDs and LBDs, respectively. For each source, the NIRCam/F115W image (corresponds to rest UV) and the RGB composite image are shown. The RGB images are built with NIRCam/F115W (blue), F277W (green), and F444W (red) when available. Note that even LBDs sometimes appear red in the RGB false color images because of the emission line excess in the F444W filter.

\begin{deluxetable}{ccccccccc}
    \label{tab:sample_lrds}
    \tablecaption{LRDs and LBDs used in this work.
  	}
    \tablewidth{0pt}
    \tablehead{
    \colhead{Root} & \colhead{PID} & \colhead{ID} & \colhead{R.A.} & \colhead{Decl.} & \colhead{$z_{\rm spec}$} & \colhead{$\nu L_{\nu, {\rm UV}}$\tablenotemark{$\dagger$}} & \colhead{$L_{ {\rm H}\alpha, {\rm broad}}$\tablenotemark{$\dagger$}} & \colhead{$L_{ {\rm H} \alpha, {\rm narrow}}$\tablenotemark{$\dagger$}} \\
    \colhead{} & \colhead{} & \colhead{} & \colhead{deg} & \colhead{deg} & \colhead{} & \colhead{$10^{43}$ erg/s} & \colhead{$10^{42}$ erg/s} & \colhead{$10^{42}$ erg/s}\\
    \colhead{(1)} & \colhead{(2)}& \colhead{(3)} & \colhead{(4)} & \colhead{(5)} & \colhead{(6)} & \colhead{(7)} & \colhead{(8)} & \colhead{(9)}
    }
    \startdata
    \multicolumn{9}{c}{LRDs}\\
    \hline
    capers-cos19 & 6368 & 4771 & 150.161029 & 2.465804 & $5.9280_{-0.0002}^{+0.0002}$ & $3.64 \pm 0.90$ & $4.15_{-0.15}^{+0.13}$ & $3.52_{-0.13}^{+0.15}$ \\
    rubies-egs53 & 4233 & 42046 & 214.795368 & 52.788847 & $5.2864_{-0.0002}^{+0.0001}$ & $144.61 \pm 21.82$ & $63.16_{-1.45}^{+1.55}$ & $32.11_{-1.54}^{+1.68}$ \\
    gto-wide-uds11 & 1215 & 4994 & 34.471129 & -5.190433 & $5.5137_{-0.0002}^{+0.0002}$ & $11.49 \pm 1.47$ & $12.23_{-0.37}^{+0.32}$ & $9.45_{-0.31}^{+0.39}$ \\
    rubies-egs61 & 4233 & 55604 & 214.983026 & 52.956001 & $6.9898_{-0.0002}^{+0.0002}$ & $73.57 \pm 9.28$ & $92.91_{-1.04}^{+1.12}$ & $21.01_{-1.02}^{+0.98}$ \\
    jades-gds06 & 1286 & 159717 & 53.097528 & -27.901260 & $5.0772_{-0.0002}^{+0.0002}$ & $9.14 \pm 1.12$ & $10.18_{-0.20}^{+0.20}$ & $4.72_{-0.19}^{+0.21}$ \\
    rubies-egs63 & 4233 & 49140 & 214.892248 & 52.877410 & $6.6885_{-0.0002}^{+0.0002}$ & $92.07 \pm 8.77$ & $114.46_{-1.00}^{+0.93}$ & $4.64_{-0.85}^{+0.92}$ \\
    cosmos-transients & 6585 & 61234 & 150.106900 & 2.360046 & $7.0016_{-0.0001}^{+0.0001}$ & $24.48 \pm 1.56$ & $14.74_{-0.24}^{+0.23}$ & $8.61_{-0.20}^{+0.24}$ \\
    jades-gdn198 & 1181 & 68797 & 189.229137 & 62.146190 & $5.0475_{-0.0001}^{+0.0001}$ & $70.30 \pm 7.75$ & $87.85_{-0.63}^{+0.55}$ & $56.60_{-0.60}^{+0.66}$ \\
    gds-barrufet-s67 & 2198 & 12577 & 53.048455 & -27.815141 & $5.2402_{-0.0003}^{+0.0003}$ & $8.08 \pm 5.66$ & $4.18_{-0.76}^{+0.50}$ & $5.05_{-0.51}^{+0.78}$ \\
    rubies-uds23 & 4233 & 172350 & 34.368951 & -5.103941 & $5.5870_{-0.0003}^{+0.0003}$ & $18.85 \pm 4.98$ & $8.52_{-0.56}^{+0.51}$ & $6.51_{-0.54}^{+0.54}$ \\
    uncover-61 & 2561 & 24968 & 3.620607 & -30.399951 & $6.3457_{-0.0002}^{+0.0002}$ & $6.22 \pm 0.66$ & $4.14_{-0.13}^{+0.12}$ & $3.39_{-0.11}^{+0.13}$ \\
    macs1149 & 1208 & 5105197 & 177.405317 & 22.377898 & $5.6844_{-0.0005}^{+0.0005}$ & $5.23 \pm 0.63$ & $1.84_{-0.38}^{+0.31}$ & $1.94_{-0.32}^{+0.40}$ \\
    rubies-egs51 & 4233 & 925921 & 215.075054 & 52.943828 & $6.9494_{-0.0005}^{+0.0005}$ & $20.39 \pm 6.06$ & $17.41_{-0.77}^{+0.69}$ & $6.01_{-0.64}^{+0.75}$ \\
    nexus-obs3 & 5105 & 12143 & 268.428339 & 65.173502 & $5.8693_{-0.0005}^{+0.0006}$ & $7.89 \pm 2.12$ & $12.68_{-0.68}^{+0.75}$ & $3.25_{-0.74}^{+0.66}$ \\
    ceers-ddt & 2750 & 1768 & 214.925759 & 52.945661 & $5.0885_{-0.0003}^{+0.0003}$ & $2.84 \pm 0.53$ & $2.35_{-0.15}^{+0.18}$ & $0.96_{-0.19}^{+0.16}$ \\
    rubies-egs51 & 4233 & 926125 & 215.137081 & 52.988554 & $5.2888_{-0.0003}^{+0.0003}$ & $10.01 \pm 2.36$ & $4.80_{-0.40}^{+0.42}$ & $2.70_{-0.43}^{+0.41}$ \\
    jades-gds05 & 1286 & 204851 & 53.138593 & -27.790253 & $5.4870_{-0.0003}^{+0.0003}$ & $13.82 \pm 1.78$ & $5.04_{-0.27}^{+0.23}$ & $4.70_{-0.23}^{+0.28}$ \\
    nexus-obs5 & 5105 & 7712 & 268.568557 & 65.174343 & $5.8748_{-0.0004}^{+0.0004}$ & $67.19 \pm 15.23$ & $36.22_{-1.28}^{+1.19}$ & $10.45_{-1.23}^{+1.24}$ \\
    rubies-egs62 & 4233 & 948917 & 214.892479 & 52.856890 & $6.7328_{-0.0004}^{+0.0004}$ & $25.60 \pm 2.99$ & $6.20_{-0.72}^{+0.59}$ & $7.22_{-0.58}^{+0.75}$ \\
    rubies-egs61 & 4233 & 61496 & 214.972441 & 52.962192 & $5.0847_{-0.0006}^{+0.0007}$ & $3.94 \pm 3.73$ & $2.69_{-0.66}^{+0.97}$ & $1.28_{-1.01}^{+0.65}$ \\
    rubies-uds3 & 4233 & 47509 & 34.264602 & -5.232586 & $5.6746_{-0.0003}^{+0.0003}$ & $19.26 \pm 1.95$ & $5.10_{-0.55}^{+0.60}$ & $5.09_{-0.65}^{+0.58}$ \\
    jades-gdn2 & 1181 & 954 & 189.151966 & 62.259635 & $6.7650_{-0.0002}^{+0.0003}$ & $14.64 \pm 1.01$ & $9.11_{-0.38}^{+0.33}$ & $5.05_{-0.33}^{+0.39}$ \\
    ceers & 1345 & 746 & 214.809142 & 52.868484 & $5.6311_{-0.0008}^{+0.0007}$ & $1.66 \pm 1.04$ & $3.33_{-0.14}^{+0.25}$ & $0.32_{-4.99}^{+0.23}$ \\
    macs1423 & 1208 & 4103248 & 215.961216 & 24.058024 & $5.7755_{-0.0003}^{+0.0003}$ & $10.42 \pm 2.06$ & $2.85_{-0.54}^{+0.43}$ & $4.66_{-0.47}^{+0.57}$ \\
    uncover-61 & 2561 & 21547 & 3.550838 & -30.406598 & $5.0574_{-0.0001}^{+0.0001}$ & $26.54 \pm 1.67$ & $4.18_{-0.26}^{+0.26}$ & $8.67_{-0.27}^{+0.29}$ \\
    uncover & 2561 & 41225 & 3.533996 & -30.353308 & $6.7723_{-0.0003}^{+0.0004}$ & $16.75 \pm 2.11$ & $3.56_{-0.37}^{+0.38}$ & $1.81_{-0.39}^{+0.39}$ \\
    capers-egs61 & 6368 & 23961 & 214.899699 & 52.812840 & $5.0019_{-0.0004}^{+0.0003}$ & $5.68 \pm 1.53$ & $1.65_{-0.40}^{+0.27}$ & $2.48_{-0.26}^{+0.42}$ \\
    \hline
    \multicolumn{9}{c}{LBDs}\\
    \hline
    abell370 & 1208 & 2112807 & 39.983828 & -1.573664 & $5.9712_{-0.0002}^{+0.0002}$ & $6.42 \pm 1.68$ & $1.64_{-0.24}^{+0.24}$ & $4.72_{-0.23}^{+0.26}$ \\
    uncover & 2561 & 8943 & 3.614087 & -30.410448 & $6.3259_{-0.0002}^{+0.0002}$ & $16.00 \pm 0.59$ & $1.95_{-0.75}^{+0.62}$ & $2.18_{-0.66}^{+0.73}$ \\
    uncover & 2561 & 23608 & 3.542813 & -30.380646 & $5.7949_{-0.0001}^{+0.0001}$ & $34.88 \pm 2.25$ & $3.80_{-0.51}^{+0.42}$ & $8.65_{-0.45}^{+0.54}$ \\
    rubies-egs53 & 4233 & 50052 & 214.823454 & 52.830277 & $5.2395_{-0.0002}^{+0.0002}$ & $20.06 \pm 3.37$ & $3.50_{-0.30}^{+0.28}$ & $7.32_{-0.30}^{+0.32}$ \\
    capers-egs65 & 6368 & 24425 & 214.987487 & 52.873114 & $5.2335_{-0.0002}^{+0.0002}$ & $6.88 \pm 0.75$ & $1.02_{-0.16}^{+0.15}$ & $1.85_{-0.16}^{+0.17}$ \\
    uncover & 2561 & 11254 & 3.580446 & -30.405023 & $6.8759_{-0.0002}^{+0.0002}$ & $10.69 \pm 0.31$ & $1.23_{-0.12}^{+0.12}$ & $2.38_{-0.12}^{+0.13}$ \\
    macsj0647 & 1433 & 1944 & 101.986107 & 70.214346 & $5.9175_{-0.0002}^{+0.0002}$ & $58.86 \pm 4.86$ & $9.38_{-0.52}^{+0.50}$ & $10.41_{-0.54}^{+0.54}$ \\
    \enddata
    \tablenotetext{}{(1) Root ID in DJA. (2) JWST/NIRSpec observation program ID. (3) Source ID in DJA. (4) Right Ascension in J2000. (5) Declination in J2000. (6) Spectroscopic redshift. (7) Rest UV continuum luminosity. (8) Broad H$\alpha$ line luminosity. (9) Narrow H$\alpha$ line luminosity.
    }
    \tablenotetext{}{$\dagger$ Luminosities are corrected for the dust attenuation (see the text).}
\end{deluxetable}

\bibliography{sample701}{}
\bibliographystyle{aasjournalv7}



\end{document}